\documentclass[ aip, jmp, amsmath,amssymb, reprint, ]{revtex4-1}
\usepackage{graphicx}% Include figure files
\usepackage{dcolumn}% Align table columns on decimal point
\usepackage{bm}% bold math
\usepackage{amssymb}
\usepackage{amsmath}
\usepackage{amsfonts}
\usepackage{comment}

\usepackage{hyperref}
\usepackage{bbm}
\usepackage[usenames]{color}
\begin{document}

\preprint{AIP/123-QED}

\title{Wavefront propagation in a bistable dual-delayed-feedback oscillator: analogy to networks with nonlocal interactions}% Force line breaks with \\
%\thanks{Footnote to title of article.}

\author{Vladimir V. Semenov}
\email{semenov.v.v.ssu@gmail.com}
\affiliation{Institute of Physics, Saratov State University, Astrakhanskaya str. 83, 410012 Saratov, Russia}

\date{\today}% It is always \today, today,
             %  but any date may be explicitly specified

\begin{abstract}
In the present research, a bistable delayed-feedback oscillator with two delayed-feedback loops is shown to replicate a network of bistable nodes with nonlocal coupling. It is demonstrated that certain aspects of the nonlocal interaction impact on wavefront propagation identified in networks of bistable elements are entirely reproduced in the dynamics of a single oscillator with two delays. In particular, adding the second delayed-feedback loop allows speeding up both deterministic and stochastic wavefront propagation, achieving stabilization of propagating fronts at lower noise intensity and preventing fronts from noise-induced destruction occurring in the presence of single delayed-feedback. All the revealed effects are studied in numerical simulations and confirmed in physical experiments, showing an excellent correspondence.
\end{abstract}

% PACS, the Physics and Astronomy
                             % Classification Scheme.
%Use showkeys class option if keyword
                              %display desired
\maketitle

\begin{quotation}
Wavefront propagation represents an interdisciplinary phenomenon observed in an incredibly broad spectrum of dynamical systems, including bistable media and ensembles of coupled bistable oscillators. In such a case, the term 'wavefront' refers to the context of fluid dynamics and describes a boundary between domains corresponding to different quiescent steady state regimes. Depending on multiple factors, the separating fronts remain to be unmoved or start to propagate such that growing spatial domains appear and either state (phase) therefore starts to dominate the whole system. A manifold of factors for wavefront propagation control includes the dynamical systems' asymmetry, the intrinsic peculiarities of internal fluctuations and external random perturbations, the properties of spatial interaction. A distinguishable class of systems exhibiting wavefront propagation are bistable delayed-feedback oscillators, which are capable of emulating the collective behaviour in networks of different topologies. Thus, certain aspects of the wavefront propagation control through adjusting the coupling properties are expected to be realized in the bistable delayed-feedback oscillators. In the current paper, this issue is discussed on an example of deterministic and stochastic wavefront propagation control in networks of nonlocally coupled bistable oscillators. Generally, the aim of the present research is to demonstrate that a single bistable dual-delayed-feedback oscillator can replicate an ensemble of nonlocally coupled bistable oscillators. This idea is confirmed from the experimental point of view, based on the revealed similarity of wavefront propagation in two kinds of dynamical systems.
\end{quotation}

\section{Introduction}
\label{intro}
The presence of long time delay in single oscillators is known to enable realization of complex phenomena that were originally found in spatially-extended systems and networks of coupled oscillators \cite{yanchuk2017}. 
Identifying spatio-temporal phenomena in the pure temporal dynamics of delayed-feedback oscillators becomes possible when applying a spatio-temporal representation, which considers the delay interval $[0,\tau]$ in analogy with the spatial coordinate  \cite{arecchi1992,giacomelli1996}.
A manifold of effects and structures revealed in the dynamics of delayed-feedback oscillators includes chimera states \cite{larger2013,larger2015,semenov2016,brunner2018}, dissipative solitons \cite{garbin2015,marconi2015,romeira2016,javaloyes2017,brunner2018,yanchuk2019,semenov2023,semenov2018} and travelling waves \cite{klinshov2017,semenov2025}, coarsening \cite{giacomelli2012} and nucleation \cite{zaks2013}. 

The capability of delay systems to replicate high-dimensional systems such as networks of interacting nodes has fundamentally impacted the hardware implementation of recurrent neural networks. In particular, delayed-feedback oscillators are leveraged for reservoir computing to implement single-node reservoirs \cite{appeltant2011,martinenghi2012,larger2017,brunner2018-2,duport2012,huelser2022,koester2022}. Moreover, time-delay reservoirs extended to deep network architectures can be implemented physically by using coupled nonlinear delayed-feedback oscillators \cite{penkovsky2019} or a single system with multiple time-delayed feedback loops \cite{stelzer2021}. Furthermore, singleton time-delay systems are promising candidates as a spin network (or spin glass) for solving problems with non-deterministic polynomial-time hardness (so-called NP-hard problems). Specifically, Ising machines based on delayed-feedback bistable systems have significant prospects for acceleration of the optimization problem computations \cite{boehm2019}.

\begin{figure}[t!]
\centering
\includegraphics[width=0.48\textwidth]{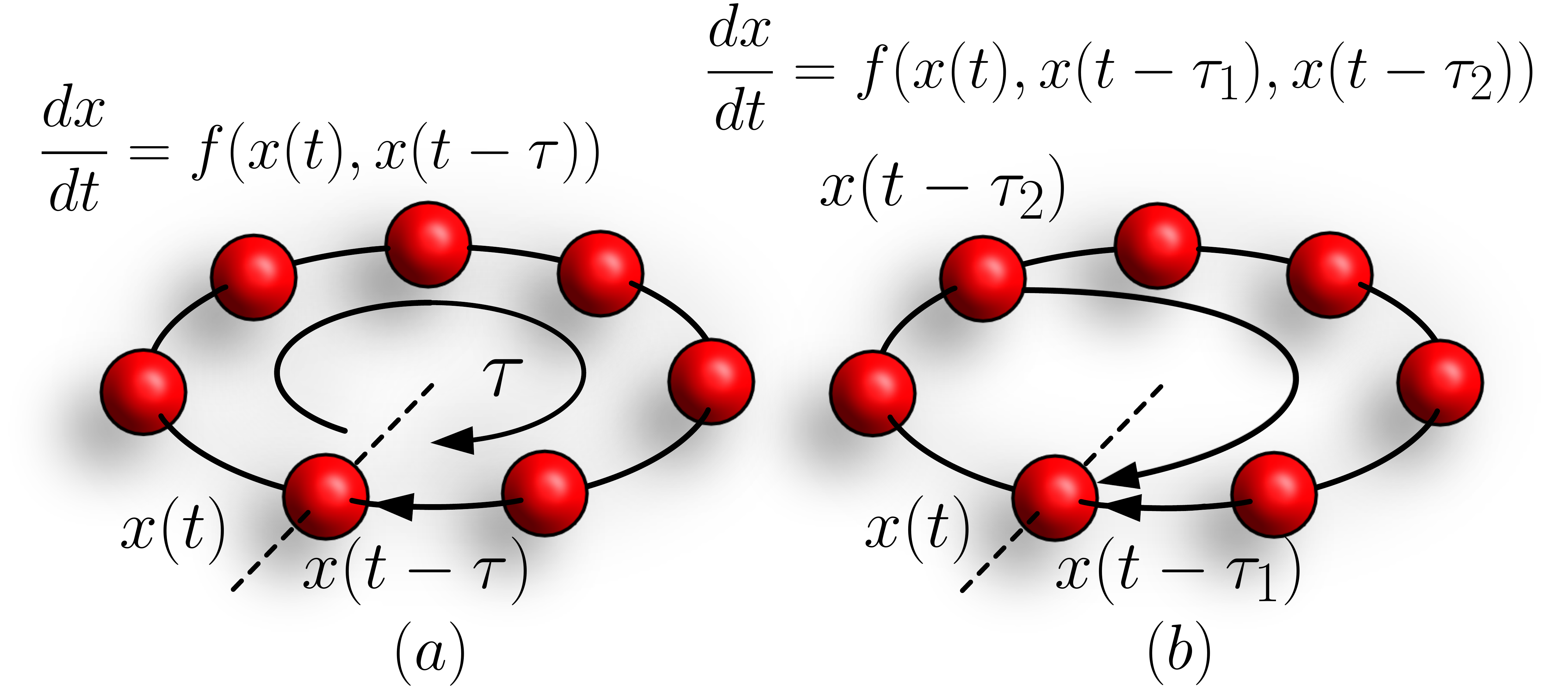}
\caption{Schematic illustration of a delayed-feedback oscillator subject to single (panel (a)) and dual (panel (b)) time-delayed feedback.}
\label{fig1}
\end{figure}

Topologically, single-delayed-feedback oscillators are equivalent to a single-layer network with unidirectional local coupling [Fig.~\ref{fig1}~(a)]. This fact is emphasized in recent paper \cite{semenov2025} on an example of the Lévy-noise-based control of wavefront propagation, where the unidirectional and bidirectional character of coupling is found to be a crucial factor. In contrast, more complicated coupling topologies are less studied in the context of replication by delayed-feedback oscillators, except for a few publications. In particular, the deep reservoir scheme considered in Ref. \cite{penkovsky2019} comprises hierarchically coupled nonlinear delay oscillators. In the context of dissipative solitons, dual-delayed-feedback oscillators are shown to replicate spatially-extended systems with nonlocal interaction \cite{javaloyes2017}. Coupled bistable delayed-feedback oscillators studied in paper \cite{zakharova2025} emulate multiplex networks, which is demonstrated on an example of wavefront propagation and stochastic resonance. In the context of this similarity, the delayed feedback strength is considered as the coupling within the layer (intra-layer coupling), whereas the coupling between two delayed-feedback oscillators corresponds to the multiplex links between the layers (inter-layer coupling).

In the present research, replication of networks with complex topology by single oscillators subject to time-delayed feedback is considered on an example of nonlocal interactions. In the context of networks of bistable oscillators, this kind of coupling is known to control wavefront propagation velocity through varying the coupling radius at fixed coupling strength and to save fronts being destroyed due to the action of noise in the presence local coupling \cite{zakharova2023,semenov2025-2}. Moreover, depending on the coupling strength value, stochastic resonance can be enhanced or suppressed when changing the coupling radius \cite{semenov2025-3}. The main idea examined in the current paper consists in the hypothesis that a single oscillator forced by two delayed-feedback loops with different delay times (schematically illustrated in Fig.~\ref{fig1}~(b)) is capable of replicating a ring with nonlocal coupling. If this statement is true, then the effects exhibited by networked oscillators and associated with the intrinsic peculiarities of nonlocal interaction must be partially or fully reproduced in the dynamics of single delayed-feedback oscillators. This issue is discussed below in the context of wavefront propagation.  

%%%%%%%%%%%%%%%%%%%%%%%% FIG 1 %%%%%%%%%%%
%%%%%%%%%%%%%%%%%%%%%%%%%%%%%%%%%%%%%%

\section{Model and methods}
\label{model_and_methods}
The phenomena discussed in the current research are explored on the example of a single bistable oscillator with two delayed-feedback loops:
\begin{equation}
\label{numerical_model}
\begin{array}{l}
\dfrac{dx}{dt}=-x(x-a)(x+b)+F(t)\\
\\
+\dfrac{\gamma}{2}  (x(t-\tau_1)+x(t-\tau_2)-2x(t)),
\end{array}
\end{equation}
where parameters $a,b>0$ define whether the oscillator's nonlinearity in the bistable regime is symmetric ($a=b$) or asymmetric ($a \neq b$), parameter $\gamma$ characterizes a delayed-feedback strength. Equation (\ref{numerical_model}) contains two time-delayed feedback terms. The first one is characterised by fixed delay time, $\tau_1=1000$ which defines the quasi-space length. The second delay time, $\tau_2$, is varied, but does not exceed the value of $\tau_1$. Two options for studying stochastic wavefront propagation in model (\ref{numerical_model}) are under consideration. In the first case,  multiplicative (parametric) noise affects the system at $a=$const and $F(t)\equiv0$ such that parameter $b$ is modulated by white Gaussian noise of intensity $D$, $b=b_0+\xi(t)$, where $\left<\xi(t)\right>=0$, $\left<\xi(t)\xi(t + \Delta t)\right> = 2D\delta(\Delta t)$. Then the variance of noise $\xi(t)$ equals to $2D$. In the second option, the impact of multiplicative noise is excluded from the consideration, $a,b=$const, whereas additive white Gaussian noise of variance $2D$ is present, $F(t)=\xi(t)$. 

\begin{figure}
\centering
\includegraphics[width=0.48\textwidth]{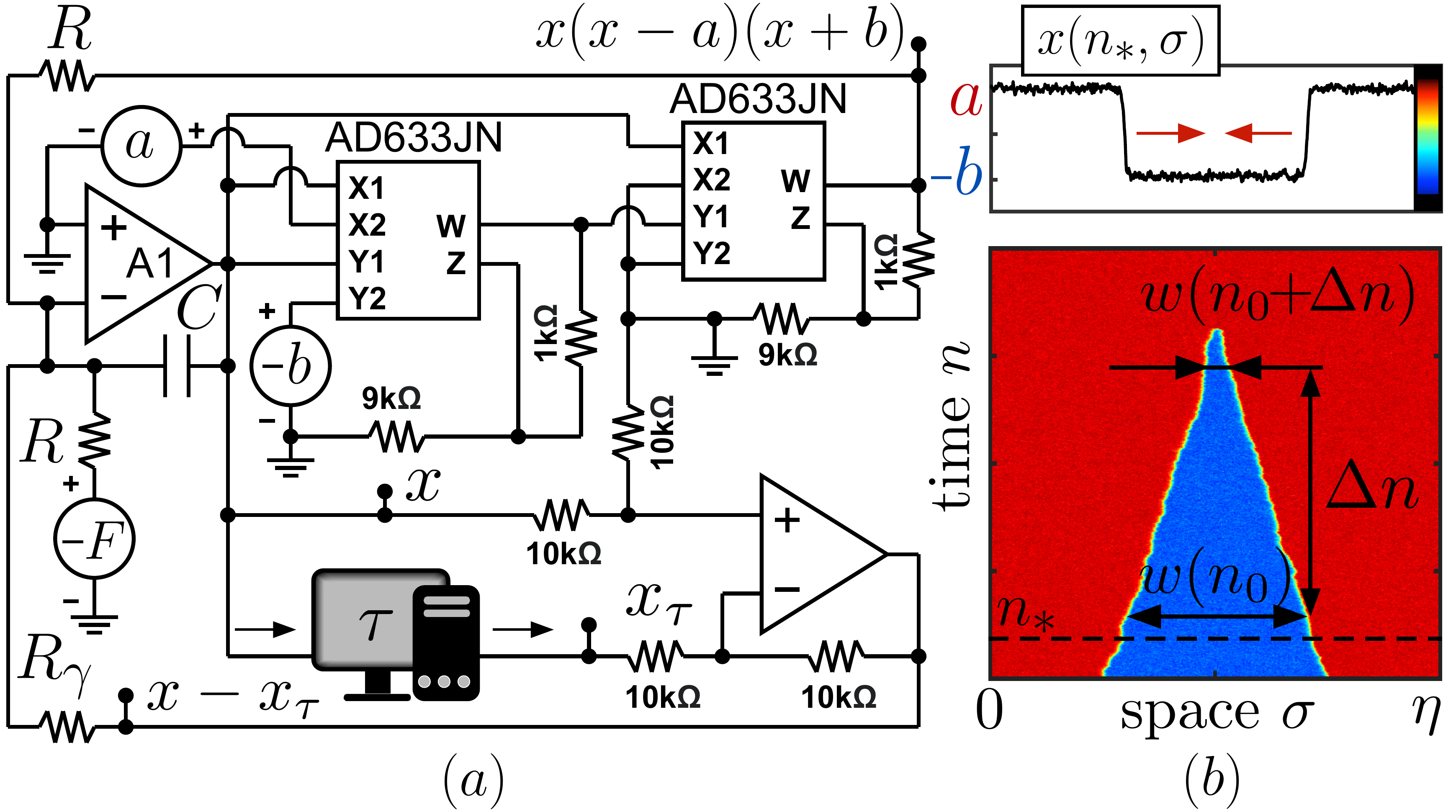} 
\caption{(a) Circuit diagram of the experimental setup (Eqs. (\ref{experimental_model})). Output voltage $x_{\tau}$ represents a sum of the delayed signals: $x_{\tau}=(x(t-\tau_1)+x(t-\tau_2))/2$. Operation amplifiers are TL072CP. Analog integrator elements are $R=10$ $k\Omega$, $R_{\gamma}=50$ $k\Omega$, $C=1$ nF; (b) Methodology for estimation of the front propagation velocity $v$ by using space-time plots. The upper inset shows the propagating fronts in quasi-space $\sigma$ at time moment $n=n_*$.}
\label{fig2}
\end{figure}

The performed investigations are carried out by means of numerical simulations and electronic experiments. In more detail, equation (\ref{numerical_model}) is integrated numerically using the Heun method \cite{mannella2002}. For physical experiments, an experimental prototype being an electronic model of system (\ref{numerical_model}) has been developed and implemented by principles of analog modelling \cite{luchinsky1998,semenov2024_book}. Certain details and solutions for the numerical and physical experiments discussed below (for instance, noise modulation of parameter values, preparing the initial conditions for the observation of wavefront propagation, etc.) are omitted in the present paper, since they have been already described in paper \cite{zakharova2025}. The circuit diagram of the setup is shown in Fig.~\ref{fig2}~(a). The central element of the electronic analog model is an inverting integrator based on an operational amplifier (see operational amplifier A1 in Fig.~\ref{fig2}~(a)). The instantaneous value of the integrator's output voltage is determined by the integral of the sum of the input signals normalized by the corresponding input resistances (see the full description in the third chapter of book \cite{semenov2024_book}). Extracting the time derivatives of of the left- and right-hand sides of the integrator equation, one can derive the mathematical model for the developed electronic device. In particular, the experimental setup in Fig.~\ref{fig2}~(a) is described by the equation
\begin{equation}
\label{experimental_model}
\begin{array}{l}
RC\dfrac{dx}{dt}=-x(x-a)(x+b)+F(t)\\
+\dfrac{\gamma}{2}  (x(t-\tau_1)+x(t-\tau_2)-2x(t)),
\end{array}
\end{equation}
where $R=10k\Omega$, $C=1$ nF, $\tau_1=69.5$ ms, $\tau_2 \leq \tau_1$, $\gamma=R/R_{\gamma}=10k\Omega/50k\Omega=0.2$. The circuit in Fig. \ref{fig2} includes a delay line which involves a personal computer as a part of the experimental setup. 
%%%%%%%%%%%%%%%%%%fig2%%%%%%%%%%%%%%%%%%%%%%
%%%%%%%%%%%%%%%%%%%%%%%%%%%%%%%%%%%%%%%%%%
The used computer is equipped by board NI PCI-6251 manufactured by National Instruments. The board contains analog-to-digital and digital-to-analog converters to acquire the input analog signal $x(t)$ and to convert it into a digital form for further processing by the PC. After the delayed signals $x(t-\tau_1)$ and $x(t-\tau_2)$ are saved as arrays in computer's memory (the sampling rate is 1 MHz), their sum divided by two is generated as analog output signal. 

The circuit contains an additive signal $F(t)=\xi(t)+F_0$ where $\xi(t)$ is responsible for additive stochastic impact and $F_0$ is a small DC-Voltage (does not exceed $\pm$5mV) which remains to be fixed during the whole experiments. The reason for including constant $F_0$ consists in the necessity to correct the additional system's asymmetry caused by individual non-zero output offset voltages of the operational amplifiers and analog multiplier. The electronic model is sensitive to such offsets. As a result, relatively slow self-sustained wavefront propagation is observed in system (\ref{experimental_model}) when parameters $a$ and $b$ are chosen to be equal and the system's nonlinearity is assumed to be symmetric. In such a case, $F_0$ is tuned to slow down wavefront propagation at $a=b$ as much as possible and to get as close to the dynamics of mathematical model (\ref{numerical_model}) as possible. A source of noise used in physical experiments, Agilent 33250A, produces broadband Gaussian noise $\xi(t)$. Similarly to numerical model, stochastic wavefront propagation is studied in experimental setup (\ref{experimental_model}) for two options: 1) $a=$const, $\xi(t)=F_0\approx 0 $, $b=b_0+\xi(t)$ (the presence of multiplicative noise); 2) $a,b=$const, $F(t)=\xi(t)+F_0$ (the impact of additive noise).

To visualize the temporal system dynamics in the quasi-space and illustrate the propagating fronts and domains, space-time plots are built. For this purpose, numerically and experimentally obtained time realizations $x(t)$ are mapped onto space-time ($\sigma,n$) by introducing $t=n\eta+\sigma$ with an integer time variable $n$, and a pseudo-space variable $\sigma \in [0,\eta]$, where $\eta=\tau_1+\varepsilon$ with a quantity $\varepsilon$, which is small as compared to $\tau_1$ and results from a finite internal response time of the system. A unique value of the quasi-space length $\eta$ is chosen in numerical and physical experiments such that space-time diagrams $x(\sigma,n)$ are vertically oriented at $\tau_2=\tau_1$ [Fig.~\ref{fig2}~(b)]. In such a case, the propagation to the left and to the right is identical as in the classical case of wavefront propagation in media and ensembles. After the appropriate value of $\eta$ is found, the wavefront propagation velocity is calculated [Fig.~\ref{fig2}~(b)]. In the current paper, the wavefront propagation is characterised by the expansion velocity of the state $a$ (the red domain). For this purpose, the propagation velocity is introduced as $v=(w(n_0)-w(n_0+\Delta n))/2 \Delta n$. Here, $w(n_0)$ and $w(n_0+\Delta n)$ are the widths of the central spatial domain  corresponding to the state $x(t)=-b$ in the quasi-space at the moments $n_0$ and $n_0+\Delta n$, respectively. The relationship for $v$ involves the propagation of two fronts, but describes the single front propagation. For this reason, the formula for $v$ includes the factor 2. In summary, quantity $v$ is the velocity of the left front propagating to the right and, similarly, the velocity of the right front moving to the left [Fig.~\ref{fig2}~(b)].

\section{Deterministic wavefront propagation}
\label{sec:deterministic}

\begin{figure}
\centering
\includegraphics[width=0.47\textwidth]{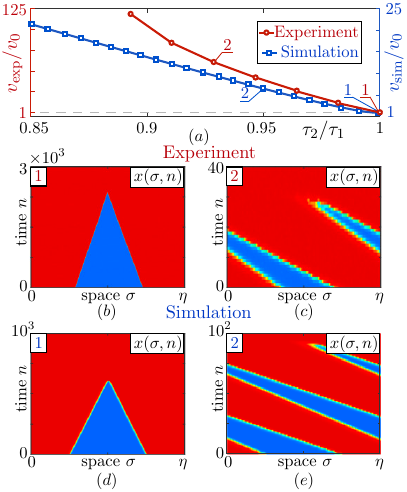} 
\caption{Deterministic wavefront propagation control in model (\ref{numerical_model}) and experimental setup (\ref{experimental_model}) thought varying $\tau_2$: (a) Dependence of the normalized wavefront propagation velocity on delay time $\tau_2$ registered in numerical simulations, $v_{\text{sim}}$, and electronic experiments, $v_{\text{exp}}$. Space-time diagrams in panels (b)-(e) illustrate the dynamics in points 1-2 in numerical and experimental dependencies on panel (a). Parameters are: $a=0.5$, $b=0.45$, $\gamma=0.2$, $\tau_1=1000$ (simulations) and $\tau_1=69.5$ ms (experiments). 
%Quantity $v_0$ is a wavefront velocity registered at $\tau_1=\tau_2$: $v_0=0.4124$ (simulations) and $v_0=6.46\times 10^{-6}$ (experiments). 
The quasi-space length is $\eta=1005.95$ (simulations) and $\eta=69.8412$ ms (experiments).}
\label{fig3}
\end{figure}

As numerically and experimentally demonstrated in Refs. \cite{semenov2025,zakharova2025}, the intrinsic peculiarities of deterministic and stochastic control of wavefront propagation revealed in single-layer and multilayer networks of locally coupled oscillators can be realized in a bistable oscillator subject to single time-delayed feedback (numerical model (\ref{numerical_model}) and experimental setup (\ref{experimental_model}) at $\tau_2=\tau_1$). When increasing the number of feedback loops, wavefront propagation is sped up. This fact is illustrated in Fig. \ref{fig3} on an example of the deterministic process (at this stage, the dynamics observed in physical experiments is also assumed to be deterministic) occurring in the asymmetric oscillator with two delayed-feedback loops where $\gamma$ and $\tau_1$ are fixed, whereas $\tau_2$ varies. In the limit case $\tau_2=\tau_1$, Eqs. (\ref{numerical_model}) and (\ref{experimental_model}) transform into the oscillator forced by single time-delayed feedback discussed in papers \cite{semenov2025,zakharova2025}. The wavefront propagation velocity measured at $\tau_2=\tau_1$ in numerical and physical experiments is used as reference value $v_0$ in Fig. \ref{fig3}: $v_0=0.4124$ (numerical modelling) and $v_0=6.46\times 10^{-6}$ (physical experiment). The difference of several orders between numerical and experimental $v_0$ results from the fact that the time scales of the mathematical model and the experimental setup differ by a factor of $RC$. In addition, the experimental setup inevitably includes additional inaccuracies and factors affecting the system's asymmetry. 

%%%%%%%%%%%%%%%%%%fig3%%%%%%%%%%%%%%%%%%%%%%
%%%%%%%%%%%%%%%%%%%%%%%%%%%%%%%%%%%%%%%%%%

When decreasing delay time $\tau_2$, the wavefront velocity significantly increases [Fig.~\ref{fig3}~(a)] and exceeds extremely high values such that the wavefronts and spatial domains disappear almost immediately after they are initialized in simulations and physical experiments (compare space-time plots in Fig. \ref{fig3}~(b)-(e) and pay attention to different scales of vertical axis). Thus, further decreasing $\tau_2$ makes no sense and case $\tau_2<0.85 \tau_1$ is excluded from consideration. Besides growth of the wavefront velocity, the presence of the second delayed-feedback loop makes the space-time plots to be tilted. In such a case, both fronts move to the left. However, this has no impact on quantity $v$ which characterises how fast the spatial domain corresponding to $x(\sigma,n)$ invades the entire available space. In addition, one can transform the space-time plots and make them vertically symmetric again by decreasing length of the quasi-space $\eta$. 

It is important to note that experimental setup (\ref{experimental_model}) is much more sensitive to decreasing $\tau_2$ as compared to mathematical model (\ref{numerical_model}): values $v/v_0$ obtained in numerical and physical experiments at the same values of $\tau_2/\tau_1$ differ by at least five times. This result correlates with materials reported in Refs. \cite{semenov2025,zakharova2025}. In the mentioned papers, the wavefront velocities registered in electronic experiments and numerical simulations at the same parameter values are shown to be different by approximately five times.

\section{Stochastic phenomena}
\label{sec:stochastic}
\begin{figure}
\centering
\includegraphics[width=0.47\textwidth]{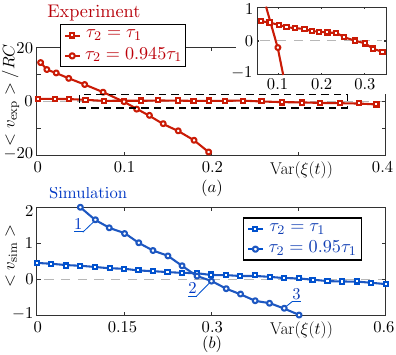} 
\caption{Multiplicative-noise-based wavefront propagation control: Dependencies of the mean wavefront propagation velocity on the variance of parametric noise ($b=0.45 + \xi(t)$) registered in electronic setup (\ref{experimental_model}) (panel (a)) and numerical model (\ref{numerical_model}) (panel (b)) at $\tau_2=\tau_1$ and $\tau_2<\tau_1$. The upper inset in panel (a) shows the experimentally obtained dependencies in ranges $<v_{\text{exp}}>/RC\in [-1,1]$ and Var$(\xi(t))\in[0.05,0.35]$ (delineated by the dashed rectangle in panel (a)).  Other parameters are: $\gamma=0.2$, $\tau_1=1000$ (simulations) and $\tau_1=69.5$ ms (experiments). The additive force is absent, $F(t)\equiv 0$.}
\label{fig4}
\end{figure}

Besides the deterministic effect, introducing the second delayed feedback also impacts the stochastic wavefront propagation. In such a case, one deals with the same rule: if the wavefront velocity is nonzero at fixed oscillator's parameters and noise intensity, the presence of second delayed feedback gives rise to speeding up the propagation. To visualise such phenomena, consider model (\ref{numerical_model}) and electronic setup (\ref{experimental_model}) in the presence of a source $\xi(t)$ of multiplicative (parametric) white Gaussian noise: $a=\text{const}=0.5$, $b=0.45+\xi(t)$, $F(t)\equiv 0$. Then increasing the noise level results first in noise-sustained stabilization (when the mean wavefront velocity is close to zero) and then to reversal of the wavefront propagation [Fig. \ref{fig4}]. However, the slope of the dependency of the mean wavefront velocity on the noise's variance is higher at $\tau_2<\tau_1$ as compared to case $\tau_2=\tau_1$. This is observed both in numerical and physical experiments. In addition, it must be noted that introducing the second delay feedback provides for achieving noise-sustained stabilization of wavefront propagation at lower values of the noise intensity. In particular, the variances of noise corresponding to almost zero mean wavefront velocity registered in numerical simulations at $\tau_1=\tau_2$ and $\tau_1=0.95\tau_2$ are Var$(\xi(t))\approx0.5$ and Var$(\xi(t))\approx0.3$. In physical experiments, the fronts are stabilized at Var$(\xi(t))\approx 0.3$ ($\tau_1=\tau_2$) and Var$(\xi(t))\approx 0.1$ ($\tau_1=0.945\tau_2$).

%%%%%%%%%%%%%%%%%%fig4%%%%%%%%%%%%%%%%%%%%%%
%%%%%%%%%%%%%%%%%%%%%%%%%%%%%%%%%%%%%%%%%%

%%%%%%%%%%%%%%%%%%fig5%%%%%%%%%%%%%%%%%%%%%%
\begin{figure}
\centering
\includegraphics[width=0.47\textwidth]{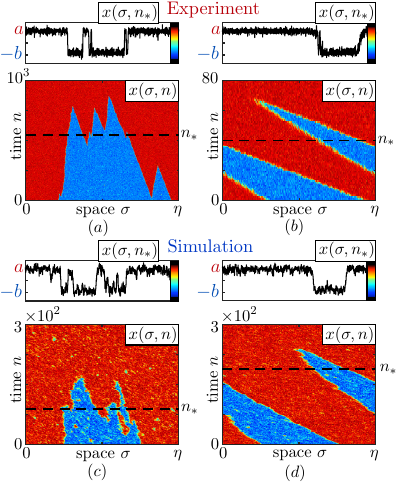} 
\caption{
Introducing the dual delayed feedback prevents wavefront propagation failure
induced by additive noise, which is observed both in physical experiments (panels (a), (b)) and numerical simulations (panels (c), (d)): the spatial domain corresponding to $x(\sigma,n)=-b$ is destroyed in case $\tau_2=\tau_1$ (panels (a),(c)) but persists when the delayed feedback becomes dual ($\tau_2=0.99\tau_1$ in panel (b) and $\tau_2=0.945\tau_1$ in panel (d)). The upper insets show the instantaneous systems' state at discrete time moments $n=n_*$ in space-time plots: $n_*=550$ (panel (a)), $n_*=40$ (panel (b)), $n_*=90$ (panel (c)), $n_*=190$ (panel (d)). Parameters of model (\ref{numerical_model}) and setup (\ref{experimental_model}) are $a=0.5$, $b=0.45$, $\gamma=0.2$, $\tau_1=1000$ (simulations) and $\tau_1=69.5$ ms (experiments). The quasi-space length is $\eta=1005.95$ (simulations) and $\eta=69.8412$ ms (experiments). Additive noise is introduced as $F(t)=\xi(t)$ where $\xi(t)$ is a source of white Gaussian noise of variance Var$(\xi(t))=0.01$ (simulations) and Var$(\xi(t))=$ (experiments). The upper insets show the instantaneous systems' state at discrete time moments $n=n_*$ in space-time plots. The quasi-space length is $\eta=1005.95$ (simulations, panels (c),(d)) and $\eta=69.8412$ ms (experiments, panels (a),(b)).}
\label{fig5}
\end{figure}  
%%%%%%%%%%%%%%%%%%%%%%%%%%%%%%%%%%%%%%%%%%

Increasing the wavefront velocity by adding extra delayed feedback and achieving the noise-sustained stabilization of wavefront propagation at lower noise intensity are interpreted in the current paper as the manifestation of the the similarity between the action of multi-delayed feedback in a single oscillator and nonlocal interaction in networks of coupled oscillators \cite{zakharova2023,semenov2025-2}. However, the reader can doubt this hypothesis. Indeed, there could be various reasons for speeding up wavefront propagation in the presence of two delays and this does not have to be a result of correspondence with nonlocal coupling. To reinforce the arguments for the similarity with nonlocal coupling, a distinguishable effect directly associated with the action of nonlocal coupling in ensembles of coupled bistable oscillators is shown below to be exhibited by a single bistable oscillator with dual-delayed feedback. In particular, it is reported in Refs. \cite{zakharova2023,semenov2025-2} that nonlocal coupling allows to save fronts which are destroyed at the same noise intensity in networks of locally-coupled bistable oscillators. 

To observe noise-induced destruction of fronts and spatial domains, consider model (\ref{numerical_model}) and experimental setup (\ref{experimental_model}) in the presence of additive noise: $a=0.5$, $b=0.45$, $F(t)=\xi(t)$, where $\xi(t)$ is a source of white Gaussian noise of variable amplitude. When increasing the variance of noise at $\tau_2=\tau_1$, the fronts inevitably begin to collapse at certain noise level [Fig.~\ref{fig5}~(a),(c)]. However, the same fronts (initialized at the same initial conditions, noise intensity and parameter values except of $\tau_2$) sustain when $\tau_2$ becomes lower than $\tau_1$ such that the state $x(\sigma,n)=a$ gradually invades the entire available space [Fig.~\ref{fig5}~(b),(d)]. Thus, adding the second time-delayed feedback prevents propagation failure in the presence of noise similarly to the impact of nonlocal coupling in networks of bistable oscillators.

\section{Conclusions}
In the current research, a hypothesis that bistable dual-delayed-feedback oscillators can reproduce the effects observed in ensembles of nonlocally coupled bistable oscillators is proposed and validated. In the context of this analogy, the delayed-feedback strength plays a role of the networks' coupling strength, whereas two delayed terms of the oscillator's equation realize two coupling links, the local and nonlocal one. The similarity of the phenomena in two systems of different classes has been analyzed from the experimental point of view. In particular, it has been demonstrated based on numerical simulations and physical experiments that the effect of wavefront propagation being typical for bistable systems is similarly manifested by networks and single oscillators with delay. 

Surprisingly, all the aspects of the nonlocal-coupling-based control of deterministic and stochastic wavefront propagation in networks of coupled oscillators reported in paper \cite{semenov2025-2} are replicated by the bistable dual-delayed-feedback oscillator with high accuracy. Particularly, adding the second delayed-feedback loop provides for speeding up the deterministic wavefront propagation naturally exhibited in the presence of asymmetry. Moreover, all the aspects of the wavefront propagation in the presence of additive and multiplicative noise also are reproduced. Firstly, the slope of the dependency of the wavefront propagation velocity on the multiplicative noise intensity increases when the second delayed-feedback loop is introduced. Secondly, in the presence of two delays, multiplicative-noise-sustained stabilization of wavefront propagation is achieved at lower noise intensity as compared to the system with a single delayed-feedback loop. Finally, adding the second delay term is found to prevent fronts from noise-induced destruction occurring in the presence of single delayed-feedback. However, the revealed difference must be noted. In contrast to the space-time plots presented in Ref. \cite{semenov2025-2}, the space-time diagrams of the delayed-feedback oscillator are tilted such that two fronts move to the left with different velocities. This is due to the fact that the oscillators considered in publication Ref. \cite{semenov2025-2} are coupled bidirectionally, whereas delayed-feedback oscillators are equivalent to a ring with unidirectional coupling. 

%Nevertheless, the chosen metodology for calculation of wavefront velocity involves estimating widths of spatial domains and works in case of tilted space-time plots: the tilt of space-time plot can be corrected by decresing the quasi-space length, which has no noticeable impact on the obtained values of the wavefront velocity. 

\section*{DATA AVAILABILITY}
The data that support the findings of this study are available from the author upon reasonable request.

\section*{Acknowledgements}
This work was supported by the Russian Science Foundation (project No. 24-72-00054). 

\text{https://rscf.ru/project/24-72-00054/}

%%\bibliography{biblio}% Produces the bibliography via BibTeX.

\begin{thebibliography}{35}%
\makeatletter
\providecommand \@ifxundefined [1]{%
 \@ifx{#1\undefined}
}%
\providecommand \@ifnum [1]{%
 \ifnum #1\expandafter \@firstoftwo
 \else \expandafter \@secondoftwo
 \fi
}%
\providecommand \@ifx [1]{%
 \ifx #1\expandafter \@firstoftwo
 \else \expandafter \@secondoftwo
 \fi
}%
\providecommand \natexlab [1]{#1}%
\providecommand \enquote  [1]{``#1''}%
\providecommand \bibnamefont  [1]{#1}%
\providecommand \bibfnamefont [1]{#1}%
\providecommand \citenamefont [1]{#1}%
\providecommand \href@noop [0]{\@secondoftwo}%
\providecommand \href [0]{\begingroup \@sanitize@url \@href}%
\providecommand \@href[1]{\@@startlink{#1}\@@href}%
\providecommand \@@href[1]{\endgroup#1\@@endlink}%
\providecommand \@sanitize@url [0]{\catcode `\\12\catcode `\$12\catcode
  `\&12\catcode `\#12\catcode `\^12\catcode `\_12\catcode `\%12\relax}%
\providecommand \@@startlink[1]{}%
\providecommand \@@endlink[0]{}%
\providecommand \url  [0]{\begingroup\@sanitize@url \@url }%
\providecommand \@url [1]{\endgroup\@href {#1}{\urlprefix }}%
\providecommand \urlprefix  [0]{URL }%
\providecommand \Eprint [0]{\href }%
\providecommand \doibase [0]{http://dx.doi.org/}%
\providecommand \selectlanguage [0]{\@gobble}%
\providecommand \bibinfo  [0]{\@secondoftwo}%
\providecommand \bibfield  [0]{\@secondoftwo}%
\providecommand \translation [1]{[#1]}%
\providecommand \BibitemOpen [0]{}%
\providecommand \bibitemStop [0]{}%
\providecommand \bibitemNoStop [0]{.\EOS\space}%
\providecommand \EOS [0]{\spacefactor3000\relax}%
\providecommand \BibitemShut  [1]{\csname bibitem#1\endcsname}%
\let\auto@bib@innerbib\@empty
%</preamble>
\bibitem [{\citenamefont {Yanchuk}\ and\ \citenamefont
  {Giacomelli}(2017)}]{yanchuk2017}%
  \BibitemOpen
  \bibfield  {author} {\bibinfo {author} {\bibfnamefont {S.}~\bibnamefont
  {Yanchuk}}\ and\ \bibinfo {author} {\bibfnamefont {G.}~\bibnamefont
  {Giacomelli}},\ }\bibfield  {title} {\enquote {\bibinfo {title}
  {Spatio-temporal phenomena in complex systems with time delays},}\
  }\href@noop {} {\bibfield  {journal} {\bibinfo  {journal} {J. Phys. A: Math.
  Theor.}\ }\textbf {\bibinfo {volume} {50}},\ \bibinfo {pages} {103001}
  (\bibinfo {year} {2017})}\BibitemShut {NoStop}%
\bibitem [{\citenamefont {Arecchi}\ \emph {et~al.}(1992)\citenamefont
  {Arecchi}, \citenamefont {Giacomelli}, \citenamefont {Lapucci},\ and\
  \citenamefont {Meucci}}]{arecchi1992}%
  \BibitemOpen
  \bibfield  {author} {\bibinfo {author} {\bibfnamefont {F.~T.}\ \bibnamefont
  {Arecchi}}, \bibinfo {author} {\bibfnamefont {G.}~\bibnamefont {Giacomelli}},
  \bibinfo {author} {\bibfnamefont {A.}~\bibnamefont {Lapucci}}, \ and\
  \bibinfo {author} {\bibfnamefont {R.}~\bibnamefont {Meucci}},\ }\bibfield
  {title} {\enquote {\bibinfo {title} {Two-dimensional representation of a
  delayed dynamical system},}\ }\href@noop {} {\bibfield  {journal} {\bibinfo
  {journal} {Phys. Rev. A}\ }\textbf {\bibinfo {volume} {45}},\ \bibinfo
  {pages} {R4225} (\bibinfo {year} {1992})}\BibitemShut {NoStop}%
\bibitem [{\citenamefont {Giacomelli}\ and\ \citenamefont
  {Politi}(1996)}]{giacomelli1996}%
  \BibitemOpen
  \bibfield  {author} {\bibinfo {author} {\bibfnamefont {G.}~\bibnamefont
  {Giacomelli}}\ and\ \bibinfo {author} {\bibfnamefont {A.}~\bibnamefont
  {Politi}},\ }\bibfield  {title} {\enquote {\bibinfo {title} {Relationship
  between delayed and spatially extended dynamical systems},}\ }\href@noop {}
  {\bibfield  {journal} {\bibinfo  {journal} {Phys. Rev. Lett.}\ }\textbf
  {\bibinfo {volume} {76}},\ \bibinfo {pages} {2686} (\bibinfo {year}
  {1996})}\BibitemShut {NoStop}%
\bibitem [{\citenamefont {Larger}, \citenamefont {Penkovsky},\ and\
  \citenamefont {Maistrenko}(2013)}]{larger2013}%
  \BibitemOpen
  \bibfield  {author} {\bibinfo {author} {\bibfnamefont {L.}~\bibnamefont
  {Larger}}, \bibinfo {author} {\bibfnamefont {B.}~\bibnamefont {Penkovsky}}, \
  and\ \bibinfo {author} {\bibfnamefont {Y.}~\bibnamefont {Maistrenko}},\
  }\bibfield  {title} {\enquote {\bibinfo {title} {Virtual chimera states for
  delayed-feedback systems},}\ }\href@noop {} {\bibfield  {journal} {\bibinfo
  {journal} {Phys. Rev. Lett.}\ }\textbf {\bibinfo {volume} {5}},\ \bibinfo
  {pages} {054103} (\bibinfo {year} {2013})}\BibitemShut {NoStop}%
\bibitem [{\citenamefont {Larger}, \citenamefont {Penkovsky},\ and\
  \citenamefont {Maistrenko}(2015)}]{larger2015}%
  \BibitemOpen
  \bibfield  {author} {\bibinfo {author} {\bibfnamefont {L.}~\bibnamefont
  {Larger}}, \bibinfo {author} {\bibfnamefont {B.}~\bibnamefont {Penkovsky}}, \
  and\ \bibinfo {author} {\bibfnamefont {Y.}~\bibnamefont {Maistrenko}},\
  }\bibfield  {title} {\enquote {\bibinfo {title} {Laser chimeras as a paradigm
  for multistable patterns in complex systems},}\ }\href@noop {} {\bibfield
  {journal} {\bibinfo  {journal} {Nature Communications}\ }\textbf {\bibinfo
  {volume} {6}},\ \bibinfo {pages} {7752} (\bibinfo {year} {2015})}\BibitemShut
  {NoStop}%
\bibitem [{\citenamefont {Semenov}\ \emph {et~al.}(2016)\citenamefont
  {Semenov}, \citenamefont {Zakharova}, \citenamefont {Maistrenko},\ and\
  \citenamefont {Sch{\"o}ll}}]{semenov2016}%
  \BibitemOpen
  \bibfield  {author} {\bibinfo {author} {\bibfnamefont {V.}~\bibnamefont
  {Semenov}}, \bibinfo {author} {\bibfnamefont {A.}~\bibnamefont {Zakharova}},
  \bibinfo {author} {\bibfnamefont {Y.}~\bibnamefont {Maistrenko}}, \ and\
  \bibinfo {author} {\bibfnamefont {E.}~\bibnamefont {Sch{\"o}ll}},\ }\bibfield
   {title} {\enquote {\bibinfo {title} {Delayed-feedback chimera states: Forced
  multiclusters and stochastic resonance},}\ }\href@noop {} {\bibfield
  {journal} {\bibinfo  {journal} {Europhys. Lett.}\ }\textbf {\bibinfo {volume}
  {115}},\ \bibinfo {pages} {10005} (\bibinfo {year} {2016})}\BibitemShut
  {NoStop}%
\bibitem [{\citenamefont {Brunner}\ \emph
  {et~al.}(2018{\natexlab{a}})\citenamefont {Brunner}, \citenamefont
  {Penkovsky}, \citenamefont {Levchenko}, \citenamefont {Sch{\"o}ll},
  \citenamefont {Larger},\ and\ \citenamefont {Maistrenko}}]{brunner2018}%
  \BibitemOpen
  \bibfield  {author} {\bibinfo {author} {\bibfnamefont {D.}~\bibnamefont
  {Brunner}}, \bibinfo {author} {\bibfnamefont {B.}~\bibnamefont {Penkovsky}},
  \bibinfo {author} {\bibfnamefont {R.}~\bibnamefont {Levchenko}}, \bibinfo
  {author} {\bibfnamefont {E.}~\bibnamefont {Sch{\"o}ll}}, \bibinfo {author}
  {\bibfnamefont {L.}~\bibnamefont {Larger}}, \ and\ \bibinfo {author}
  {\bibfnamefont {Y.}~\bibnamefont {Maistrenko}},\ }\bibfield  {title}
  {\enquote {\bibinfo {title} {Two-dimensional spatiotemporal complexity in
  dual-delayed nonlinear feedback systems: Chimeras and dissipative
  solitons},}\ }\href@noop {} {\bibfield  {journal} {\bibinfo  {journal}
  {Chaos}\ }\textbf {\bibinfo {volume} {28}},\ \bibinfo {pages} {103106}
  (\bibinfo {year} {2018}{\natexlab{a}})}\BibitemShut {NoStop}%
\bibitem [{\citenamefont {Garbin}\ \emph {et~al.}(2015)\citenamefont {Garbin},
  \citenamefont {Javaloyes}, \citenamefont {Tissoni},\ and\ \citenamefont
  {Barland}}]{garbin2015}%
  \BibitemOpen
  \bibfield  {author} {\bibinfo {author} {\bibfnamefont {B.}~\bibnamefont
  {Garbin}}, \bibinfo {author} {\bibfnamefont {J.}~\bibnamefont {Javaloyes}},
  \bibinfo {author} {\bibfnamefont {G.}~\bibnamefont {Tissoni}}, \ and\
  \bibinfo {author} {\bibfnamefont {S.}~\bibnamefont {Barland}},\ }\bibfield
  {title} {\enquote {\bibinfo {title} {Topological solitons as addressable
  phase bits in a driven laser},}\ }\href@noop {} {\bibfield  {journal}
  {\bibinfo  {journal} {Nature Communications}\ }\textbf {\bibinfo {volume}
  {6}},\ \bibinfo {pages} {5915} (\bibinfo {year} {2015})}\BibitemShut
  {NoStop}%
\bibitem [{\citenamefont {Marconi}\ \emph {et~al.}(2015)\citenamefont
  {Marconi}, \citenamefont {Javaloyes}, \citenamefont {Barland}, \citenamefont
  {Balle},\ and\ \citenamefont {Giudici}}]{marconi2015}%
  \BibitemOpen
  \bibfield  {author} {\bibinfo {author} {\bibfnamefont {M.}~\bibnamefont
  {Marconi}}, \bibinfo {author} {\bibfnamefont {J.}~\bibnamefont {Javaloyes}},
  \bibinfo {author} {\bibfnamefont {S.}~\bibnamefont {Barland}}, \bibinfo
  {author} {\bibfnamefont {S.}~\bibnamefont {Balle}}, \ and\ \bibinfo {author}
  {\bibfnamefont {M.}~\bibnamefont {Giudici}},\ }\bibfield  {title} {\enquote
  {\bibinfo {title} {Vectorial dissipative solitons in vertical-cavity
  surface-emitting lasers with delays},}\ }\href@noop {} {\bibfield  {journal}
  {\bibinfo  {journal} {Nature Photonics}\ }\textbf {\bibinfo {volume} {9}},\
  \bibinfo {pages} {450} (\bibinfo {year} {2015})}\BibitemShut {NoStop}%
\bibitem [{\citenamefont {Romeira}\ \emph {et~al.}(2016)\citenamefont
  {Romeira}, \citenamefont {Av{\'o}}, \citenamefont {Figueiredo}, \citenamefont
  {Barland},\ and\ \citenamefont {Javaloyes}}]{romeira2016}%
  \BibitemOpen
  \bibfield  {author} {\bibinfo {author} {\bibfnamefont {B.}~\bibnamefont
  {Romeira}}, \bibinfo {author} {\bibfnamefont {R.}~\bibnamefont {Av{\'o}}},
  \bibinfo {author} {\bibfnamefont {J.~M.~L.}\ \bibnamefont {Figueiredo}},
  \bibinfo {author} {\bibfnamefont {S.}~\bibnamefont {Barland}}, \ and\
  \bibinfo {author} {\bibfnamefont {J.}~\bibnamefont {Javaloyes}},\ }\bibfield
  {title} {\enquote {\bibinfo {title} {Regenerative memory in time-delayed
  neuromorphic photonic resonators},}\ }\href@noop {} {\bibfield  {journal}
  {\bibinfo  {journal} {Scientific Reports}\ }\textbf {\bibinfo {volume} {6}},\
  \bibinfo {pages} {19510} (\bibinfo {year} {2016})}\BibitemShut {NoStop}%
\bibitem [{\citenamefont {Javaloyes}, \citenamefont {Marconi},\ and\
  \citenamefont {Giudici}(2017)}]{javaloyes2017}%
  \BibitemOpen
  \bibfield  {author} {\bibinfo {author} {\bibfnamefont {J.}~\bibnamefont
  {Javaloyes}}, \bibinfo {author} {\bibfnamefont {M.}~\bibnamefont {Marconi}},
  \ and\ \bibinfo {author} {\bibfnamefont {M.}~\bibnamefont {Giudici}},\
  }\bibfield  {title} {\enquote {\bibinfo {title} {Nonlocality induces chains
  of nested dissipative solitons},}\ }\href@noop {} {\bibfield  {journal}
  {\bibinfo  {journal} {Phys. Rev. Lett.}\ }\textbf {\bibinfo {volume} {119}},\
  \bibinfo {pages} {033904} (\bibinfo {year} {2017})}\BibitemShut {NoStop}%
\bibitem [{\citenamefont {Yanchuk}\ \emph {et~al.}(2019)\citenamefont
  {Yanchuk}, \citenamefont {Ruschel}, \citenamefont {Sieber},\ and\
  \citenamefont {Wolfrum}}]{yanchuk2019}%
  \BibitemOpen
  \bibfield  {author} {\bibinfo {author} {\bibfnamefont {S.}~\bibnamefont
  {Yanchuk}}, \bibinfo {author} {\bibfnamefont {S.}~\bibnamefont {Ruschel}},
  \bibinfo {author} {\bibfnamefont {J.}~\bibnamefont {Sieber}}, \ and\ \bibinfo
  {author} {\bibfnamefont {M.}~\bibnamefont {Wolfrum}},\ }\bibfield  {title}
  {\enquote {\bibinfo {title} {Temporal dissipative solitons in time-delay
  feedback systems},}\ }\href@noop {} {\bibfield  {journal} {\bibinfo
  {journal} {Phys. Rev. Lett.}\ }\textbf {\bibinfo {volume} {123}},\ \bibinfo
  {pages} {053901} (\bibinfo {year} {2019})}\BibitemShut {NoStop}%
\bibitem [{\citenamefont {Semenov}, \citenamefont {Bukh},\ and\ \citenamefont
  {Semenova}(2023)}]{semenov2023}%
  \BibitemOpen
  \bibfield  {author} {\bibinfo {author} {\bibfnamefont {V.}~\bibnamefont
  {Semenov}}, \bibinfo {author} {\bibfnamefont {A.}~\bibnamefont {Bukh}}, \
  and\ \bibinfo {author} {\bibfnamefont {N.}~\bibnamefont {Semenova}},\
  }\bibfield  {title} {\enquote {\bibinfo {title} {Delay-induced
  self-oscillation excitation in the FitzHugh-Nagumo model: regular and chaotic
  dynamics},}\ }\href@noop {} {\bibfield  {journal} {\bibinfo  {journal}
  {Chaos, Solitons and Fractals}\ }\textbf {\bibinfo {volume} {172}},\ \bibinfo
  {pages} {113524} (\bibinfo {year} {2023})}\BibitemShut {NoStop}%
\bibitem [{\citenamefont {Semenov}\ and\ \citenamefont
  {Maistrenko}(2018)}]{semenov2018}%
  \BibitemOpen
  \bibfield  {author} {\bibinfo {author} {\bibfnamefont {V.}~\bibnamefont
  {Semenov}}\ and\ \bibinfo {author} {\bibfnamefont {Y.}~\bibnamefont
  {Maistrenko}},\ }\bibfield  {title} {\enquote {\bibinfo {title} {Dissipative
  solitons for bistable delayed-feedback systems},}\ }\href@noop {} {\bibfield
  {journal} {\bibinfo  {journal} {Chaos}\ }\textbf {\bibinfo {volume} {28}},\
  \bibinfo {pages} {101103} (\bibinfo {year} {2018})}\BibitemShut {NoStop}%
\bibitem [{\citenamefont {Klinshov}\ \emph {et~al.}(2017)\citenamefont
  {Klinshov}, \citenamefont {Shchapin}, \citenamefont {Yanchuk}, \citenamefont
  {Wolfrum}, \citenamefont {D'Huys},\ and\ \citenamefont
  {Nekorkin}}]{klinshov2017}%
  \BibitemOpen
  \bibfield  {author} {\bibinfo {author} {\bibfnamefont {V.}~\bibnamefont
  {Klinshov}}, \bibinfo {author} {\bibfnamefont {D.}~\bibnamefont {Shchapin}},
  \bibinfo {author} {\bibfnamefont {S.}~\bibnamefont {Yanchuk}}, \bibinfo
  {author} {\bibfnamefont {M.}~\bibnamefont {Wolfrum}}, \bibinfo {author}
  {\bibfnamefont {O.}~\bibnamefont {D'Huys}}, \ and\ \bibinfo {author}
  {\bibfnamefont {V.}~\bibnamefont {Nekorkin}},\ }\bibfield  {title} {\enquote
  {\bibinfo {title} {Embedding the dynamics of a single delay system into a
  feed-forward ring},}\ }\href@noop {} {\bibfield  {journal} {\bibinfo
  {journal} {Phys. Rev. E}\ }\textbf {\bibinfo {volume} {96}},\ \bibinfo
  {pages} {042217} (\bibinfo {year} {2017})}\BibitemShut {NoStop}%
\bibitem [{\citenamefont {Semenov}(2025{\natexlab{a}})}]{semenov2025}%
  \BibitemOpen
  \bibfield  {author} {\bibinfo {author} {\bibfnamefont {V.}~\bibnamefont
  {Semenov}},\ }\bibfield  {title} {\enquote {\bibinfo {title}
  {L{\'e}vy-noise-induced wavefront propagation for bistable systems},}\
  }\href@noop {} {\bibfield  {journal} {\bibinfo  {journal} {Chaos, Solitons
  and Fractals}\ }\textbf {\bibinfo {volume} {198}},\ \bibinfo {pages} {116533}
  (\bibinfo {year} {2025}{\natexlab{a}})}\BibitemShut {NoStop}%
\bibitem [{\citenamefont {Giacomelli}\ \emph {et~al.}(2012)\citenamefont
  {Giacomelli}, \citenamefont {Marino}, \citenamefont {Zaks},\ and\
  \citenamefont {Yanchuk}}]{giacomelli2012}%
  \BibitemOpen
  \bibfield  {author} {\bibinfo {author} {\bibfnamefont {G.}~\bibnamefont
  {Giacomelli}}, \bibinfo {author} {\bibfnamefont {F.}~\bibnamefont {Marino}},
  \bibinfo {author} {\bibfnamefont {M.~A.}\ \bibnamefont {Zaks}}, \ and\
  \bibinfo {author} {\bibfnamefont {S.}~\bibnamefont {Yanchuk}},\ }\bibfield
  {title} {\enquote {\bibinfo {title} {Coarsening in a bistable system with
  long-delayed feedback},}\ }\href@noop {} {\bibfield  {journal} {\bibinfo
  {journal} {Europhys. Lett.}\ }\textbf {\bibinfo {volume} {99}},\ \bibinfo
  {pages} {58005} (\bibinfo {year} {2012})}\BibitemShut {NoStop}%
\bibitem [{\citenamefont {Giacomelli}\ \emph {et~al.}(2013)\citenamefont
  {Giacomelli}, \citenamefont {Marino}, \citenamefont {Zaks},\ and\
  \citenamefont {Yanchuk}}]{zaks2013}%
  \BibitemOpen
  \bibfield  {author} {\bibinfo {author} {\bibfnamefont {G.}~\bibnamefont
  {Giacomelli}}, \bibinfo {author} {\bibfnamefont {F.}~\bibnamefont {Marino}},
  \bibinfo {author} {\bibfnamefont {M.}~\bibnamefont {Zaks}}, \ and\ \bibinfo
  {author} {\bibfnamefont {S.}~\bibnamefont {Yanchuk}},\ }\bibfield  {title}
  {\enquote {\bibinfo {title} {Nucleation in bistable dynamical systems with
  long delay},}\ }\href@noop {} {\bibfield  {journal} {\bibinfo  {journal}
  {Phys. Rev. E}\ }\textbf {\bibinfo {volume} {88}},\ \bibinfo {pages} {062920}
  (\bibinfo {year} {2013})}\BibitemShut {NoStop}%
\bibitem [{\citenamefont {Appeltant}\ \emph {et~al.}(2011)\citenamefont
  {Appeltant}, \citenamefont {Soriano}, \citenamefont {Van~der Sande},
  \citenamefont {Danckaert}, \citenamefont {Massar}, \citenamefont {Dambre},
  \citenamefont {Schrauwen}, \citenamefont {Mirasso},\ and\ \citenamefont
  {Fischer}}]{appeltant2011}%
  \BibitemOpen
  \bibfield  {author} {\bibinfo {author} {\bibfnamefont {L.}~\bibnamefont
  {Appeltant}}, \bibinfo {author} {\bibfnamefont {M.}~\bibnamefont {Soriano}},
  \bibinfo {author} {\bibfnamefont {G.}~\bibnamefont {Van~der Sande}}, \bibinfo
  {author} {\bibfnamefont {J.}~\bibnamefont {Danckaert}}, \bibinfo {author}
  {\bibfnamefont {S.}~\bibnamefont {Massar}}, \bibinfo {author} {\bibfnamefont
  {J.}~\bibnamefont {Dambre}}, \bibinfo {author} {\bibfnamefont
  {B.}~\bibnamefont {Schrauwen}}, \bibinfo {author} {\bibfnamefont
  {C.}~\bibnamefont {Mirasso}}, \ and\ \bibinfo {author} {\bibfnamefont
  {I.}~\bibnamefont {Fischer}},\ }\bibfield  {title} {\enquote {\bibinfo
  {title} {Information processing using a single dynamical node as complex
  system},}\ }\href@noop {} {\bibfield  {journal} {\bibinfo  {journal} {Nature
  Communications}\ }\textbf {\bibinfo {volume} {2}},\ \bibinfo {pages} {468}
  (\bibinfo {year} {2011})}\BibitemShut {NoStop}%
\bibitem [{\citenamefont {Martinenghi}\ \emph {et~al.}(2012)\citenamefont
  {Martinenghi}, \citenamefont {Rybalko}, \citenamefont {Jacquot},
  \citenamefont {Chembo},\ and\ \citenamefont {Larger}}]{martinenghi2012}%
  \BibitemOpen
  \bibfield  {author} {\bibinfo {author} {\bibfnamefont {R.}~\bibnamefont
  {Martinenghi}}, \bibinfo {author} {\bibfnamefont {S.}~\bibnamefont
  {Rybalko}}, \bibinfo {author} {\bibfnamefont {M.}~\bibnamefont {Jacquot}},
  \bibinfo {author} {\bibfnamefont {Y.~K.}\ \bibnamefont {Chembo}}, \ and\
  \bibinfo {author} {\bibfnamefont {L.}~\bibnamefont {Larger}},\ }\bibfield
  {title} {\enquote {\bibinfo {title} {Photonic nonlinear transient computing
  with multiple-delay wavelength dynamics},}\ }\href@noop {} {\bibfield
  {journal} {\bibinfo  {journal} {Phys. Rev. Lett.}\ }\textbf {\bibinfo
  {volume} {108}},\ \bibinfo {pages} {244101} (\bibinfo {year}
  {2012})}\BibitemShut {NoStop}%
\bibitem [{\citenamefont {Larger}\ \emph {et~al.}(2017)\citenamefont {Larger},
  \citenamefont {Bayl{\'o}n-Fuentes}, \citenamefont {Martinenghi},
  \citenamefont {Udaltsov}, \citenamefont {Chembo},\ and\ \citenamefont
  {Jacquot}}]{larger2017}%
  \BibitemOpen
  \bibfield  {author} {\bibinfo {author} {\bibfnamefont {L.}~\bibnamefont
  {Larger}}, \bibinfo {author} {\bibfnamefont {A.}~\bibnamefont
  {Bayl{\'o}n-Fuentes}}, \bibinfo {author} {\bibfnamefont {R.}~\bibnamefont
  {Martinenghi}}, \bibinfo {author} {\bibfnamefont {V.~S.}\ \bibnamefont
  {Udaltsov}}, \bibinfo {author} {\bibfnamefont {Y.~K.}\ \bibnamefont
  {Chembo}}, \ and\ \bibinfo {author} {\bibfnamefont {M.}~\bibnamefont
  {Jacquot}},\ }\bibfield  {title} {\enquote {\bibinfo {title} {High-speed
  photonic reservoir computing using a time-delay-based architecture: Million
  words per second classification},}\ }\href@noop {} {\bibfield  {journal}
  {\bibinfo  {journal} {Phys. Rev. X}\ }\textbf {\bibinfo {volume} {7}},\
  \bibinfo {pages} {011015} (\bibinfo {year} {2017})}\BibitemShut {NoStop}%
\bibitem [{\citenamefont {Brunner}\ \emph
  {et~al.}(2018{\natexlab{b}})\citenamefont {Brunner}, \citenamefont
  {Penkovsky}, \citenamefont {Marquez}, \citenamefont {Jacquot}, \citenamefont
  {Fisher},\ and\ \citenamefont {Larger}}]{brunner2018-2}%
  \BibitemOpen
  \bibfield  {author} {\bibinfo {author} {\bibfnamefont {D.}~\bibnamefont
  {Brunner}}, \bibinfo {author} {\bibfnamefont {B.}~\bibnamefont {Penkovsky}},
  \bibinfo {author} {\bibfnamefont {B.}~\bibnamefont {Marquez}}, \bibinfo
  {author} {\bibfnamefont {M.}~\bibnamefont {Jacquot}}, \bibinfo {author}
  {\bibfnamefont {I.}~\bibnamefont {Fisher}}, \ and\ \bibinfo {author}
  {\bibfnamefont {L.}~\bibnamefont {Larger}},\ }\bibfield  {title} {\enquote
  {\bibinfo {title} {Tutorial: Photonic neural networks in delay systems},}\
  }\href@noop {} {\bibfield  {journal} {\bibinfo  {journal} {J. Appl. Phys.}\
  }\textbf {\bibinfo {volume} {124}},\ \bibinfo {pages} {152004} (\bibinfo
  {year} {2018}{\natexlab{b}})}\BibitemShut {NoStop}%
\bibitem [{\citenamefont {Duport}\ \emph {et~al.}(2012)\citenamefont {Duport},
  \citenamefont {Schneider}, \citenamefont {Smerieri}, \citenamefont
  {Haelterman},\ and\ \citenamefont {Massar}}]{duport2012}%
  \BibitemOpen
  \bibfield  {author} {\bibinfo {author} {\bibfnamefont {F.}~\bibnamefont
  {Duport}}, \bibinfo {author} {\bibfnamefont {B.}~\bibnamefont {Schneider}},
  \bibinfo {author} {\bibfnamefont {A.}~\bibnamefont {Smerieri}}, \bibinfo
  {author} {\bibfnamefont {M.}~\bibnamefont {Haelterman}}, \ and\ \bibinfo
  {author} {\bibfnamefont {S.}~\bibnamefont {Massar}},\ }\bibfield  {title}
  {\enquote {\bibinfo {title} {All-optical reservoir computing},}\ }\href@noop
  {} {\bibfield  {journal} {\bibinfo  {journal} {Optics Express}\ }\textbf
  {\bibinfo {volume} {20}},\ \bibinfo {pages} {22783--22795} (\bibinfo {year}
  {2012})}\BibitemShut {NoStop}%
\bibitem [{\citenamefont {H{\"u}lser}\ \emph {et~al.}(2022)\citenamefont
  {H{\"u}lser}, \citenamefont {K{\"o}ster}, \citenamefont {Jaurigue},\ and\
  \citenamefont {L{\"u}dge}}]{huelser2022}%
  \BibitemOpen
  \bibfield  {author} {\bibinfo {author} {\bibfnamefont {T.}~\bibnamefont
  {H{\"u}lser}}, \bibinfo {author} {\bibfnamefont {F.}~\bibnamefont
  {K{\"o}ster}}, \bibinfo {author} {\bibfnamefont {L.}~\bibnamefont
  {Jaurigue}}, \ and\ \bibinfo {author} {\bibfnamefont {K.}~\bibnamefont
  {L{\"u}dge}},\ }\bibfield  {title} {\enquote {\bibinfo {title} {Role of
  delay-times in delay-based photonic reservoir computing},}\ }\href@noop {}
  {\bibfield  {journal} {\bibinfo  {journal} {Optical Materials Express}\
  }\textbf {\bibinfo {volume} {12}},\ \bibinfo {pages} {1214--1231} (\bibinfo
  {year} {2022})}\BibitemShut {NoStop}%
\bibitem [{\citenamefont {K{\"o}ster}, \citenamefont {Yanchuk},\ and\
  \citenamefont {L{\"u}dge}(2024)}]{koester2022}%
  \BibitemOpen
  \bibfield  {author} {\bibinfo {author} {\bibfnamefont {F.}~\bibnamefont
  {K{\"o}ster}}, \bibinfo {author} {\bibfnamefont {S.}~\bibnamefont {Yanchuk}},
  \ and\ \bibinfo {author} {\bibfnamefont {K.}~\bibnamefont {L{\"u}dge}},\
  }\bibfield  {title} {\enquote {\bibinfo {title} {Master memory function for
  delay-based reservoir computers with single-variable dynamics},}\ }\href@noop
  {} {\bibfield  {journal} {\bibinfo  {journal} {IEEE Transactions on Neural
  Networks and Learning Systems}\ }\textbf {\bibinfo {volume} {35}},\ \bibinfo
  {pages} {7712--7725} (\bibinfo {year} {2024})}\BibitemShut {NoStop}%
\bibitem [{\citenamefont {Penkovsky}\ \emph {et~al.}(2019)\citenamefont
  {Penkovsky}, \citenamefont {Porte}, \citenamefont {Jacquot}, \citenamefont
  {Larger},\ and\ \citenamefont {Brunner}}]{penkovsky2019}%
  \BibitemOpen
  \bibfield  {author} {\bibinfo {author} {\bibfnamefont {B.}~\bibnamefont
  {Penkovsky}}, \bibinfo {author} {\bibfnamefont {X.}~\bibnamefont {Porte}},
  \bibinfo {author} {\bibfnamefont {M.}~\bibnamefont {Jacquot}}, \bibinfo
  {author} {\bibfnamefont {L.}~\bibnamefont {Larger}}, \ and\ \bibinfo {author}
  {\bibfnamefont {D.}~\bibnamefont {Brunner}},\ }\bibfield  {title} {\enquote
  {\bibinfo {title} {Coupled nonlinear delay systems as deep convolutional
  neural networks},}\ }\href@noop {} {\bibfield  {journal} {\bibinfo  {journal}
  {Phys. Rev. Lett.}\ }\textbf {\bibinfo {volume} {123}},\ \bibinfo {pages}
  {054101} (\bibinfo {year} {2019})}\BibitemShut {NoStop}%
\bibitem [{\citenamefont {Stelzer}\ \emph {et~al.}(2021)\citenamefont
  {Stelzer}, \citenamefont {R{\"o}hm}, \citenamefont {Vicente}, \citenamefont
  {Fisher},\ and\ \citenamefont {Yanchuk}}]{stelzer2021}%
  \BibitemOpen
  \bibfield  {author} {\bibinfo {author} {\bibfnamefont {F.}~\bibnamefont
  {Stelzer}}, \bibinfo {author} {\bibfnamefont {A.}~\bibnamefont {R{\"o}hm}},
  \bibinfo {author} {\bibfnamefont {R.}~\bibnamefont {Vicente}}, \bibinfo
  {author} {\bibfnamefont {I.}~\bibnamefont {Fisher}}, \ and\ \bibinfo {author}
  {\bibfnamefont {S.}~\bibnamefont {Yanchuk}},\ }\bibfield  {title} {\enquote
  {\bibinfo {title} {Deep neural networks using a single neuron: folded-in-time
  architecture using feedback- modulated delay loops},}\ }\href@noop {}
  {\bibfield  {journal} {\bibinfo  {journal} {Nature Communications}\ }\textbf
  {\bibinfo {volume} {12}},\ \bibinfo {pages} {5164} (\bibinfo {year}
  {2021})}\BibitemShut {NoStop}%
\bibitem [{\citenamefont {B{\"o}hm}, \citenamefont {Verschaffelt},\ and\
  \citenamefont {Van~der Sande}(2019)}]{boehm2019}%
  \BibitemOpen
  \bibfield  {author} {\bibinfo {author} {\bibfnamefont {F.}~\bibnamefont
  {B{\"o}hm}}, \bibinfo {author} {\bibfnamefont {G.}~\bibnamefont
  {Verschaffelt}}, \ and\ \bibinfo {author} {\bibfnamefont {G.}~\bibnamefont
  {Van~der Sande}},\ }\bibfield  {title} {\enquote {\bibinfo {title} {A poor
  man's coherent Ising machine based on opto-electronic feedback systems for
  solving optimization problems},}\ }\href@noop {} {\bibfield  {journal}
  {\bibinfo  {journal} {Nature Communications}\ }\textbf {\bibinfo {volume}
  {10}},\ \bibinfo {pages} {3538} (\bibinfo {year} {2019})}\BibitemShut
  {NoStop}%
\bibitem [{\citenamefont {Zakharova}\ and\ \citenamefont
  {Semenov}(2025)}]{zakharova2025}%
  \BibitemOpen
  \bibfield  {author} {\bibinfo {author} {\bibfnamefont {A.}~\bibnamefont
  {Zakharova}}\ and\ \bibinfo {author} {\bibfnamefont {V.}~\bibnamefont
  {Semenov}},\ }\bibfield  {title} {\enquote {\bibinfo {title}
  {Delayed-feedback oscillators replicate the dynamics of multiplex networks:
  wavefront propagation and stochastic resonance},}\ }\href@noop {} {\bibfield
  {journal} {\bibinfo  {journal} {Neural networks}\ }\textbf {\bibinfo {volume}
  {183}},\ \bibinfo {pages} {106939} (\bibinfo {year} {2025})}\BibitemShut
  {NoStop}%
\bibitem [{\citenamefont {Zakharova}\ and\ \citenamefont
  {Semenov}(2023)}]{zakharova2023}%
  \BibitemOpen
  \bibfield  {author} {\bibinfo {author} {\bibfnamefont {A.}~\bibnamefont
  {Zakharova}}\ and\ \bibinfo {author} {\bibfnamefont {V.}~\bibnamefont
  {Semenov}},\ }\bibfield  {title} {\enquote {\bibinfo {title} {Stochastic
  control of spiking activity bump expansion: monotonic and resonant
  phenomena},}\ }\href@noop {} {\bibfield  {journal} {\bibinfo  {journal}
  {Chaos}\ }\textbf {\bibinfo {volume} {33}},\ \bibinfo {pages} {081101}
  (\bibinfo {year} {2023})}\BibitemShut {NoStop}%
\bibitem [{\citenamefont {Semenov}(2025{\natexlab{b}})}]{semenov2025-2}%
  \BibitemOpen
  \bibfield  {author} {\bibinfo {author} {\bibfnamefont {V.}~\bibnamefont
  {Semenov}},\ }\bibfield  {title} {\enquote {\bibinfo {title} {The impact of
  nonlocal coupling on deterministic and stochastic wavefront propagation in an
  ensemble of bistable oscillators},}\ }\href@noop {} {\bibfield  {journal}
  {\bibinfo  {journal} {Phys. Lett. A}\ }\textbf {\bibinfo {volume} {532}},\
  \bibinfo {pages} {130189} (\bibinfo {year} {2025}{\natexlab{b}})}\BibitemShut
  {NoStop}%
\bibitem [{\citenamefont {Semenov}(2025{\natexlab{c}})}]{semenov2025-3}%
  \BibitemOpen
  \bibfield  {author} {\bibinfo {author} {\bibfnamefont {V.}~\bibnamefont
  {Semenov}},\ }\bibfield  {title} {\enquote {\bibinfo {title}
  {Nonlocal-coupling-based control of stochastic resonance},}\ }\href {\doibase
  10.1140/epjs/s11734-025-01567-2} {\bibfield  {journal} {\bibinfo  {journal}
  {Eur. Phys. J. Spec. Top.}\ } (\bibinfo {year} {2025}{\natexlab{c}}),\
  10.1140/epjs/s11734-025-01567-2}\BibitemShut {NoStop}%
\bibitem [{\citenamefont {Mannella}(2002)}]{mannella2002}%
  \BibitemOpen
  \bibfield  {author} {\bibinfo {author} {\bibfnamefont {R.}~\bibnamefont
  {Mannella}},\ }\bibfield  {title} {\enquote {\bibinfo {title} {Integration of
  stochastic differential equations on a computer},}\ }\href@noop {} {\bibfield
   {journal} {\bibinfo  {journal} {International Journal of Modern Physics C}\
  }\textbf {\bibinfo {volume} {13}},\ \bibinfo {pages} {1177--1194} (\bibinfo
  {year} {2002})}\BibitemShut {NoStop}%
\bibitem [{\citenamefont {Luchinsky}, \citenamefont {McClintock},\ and\
  \citenamefont {Dykman}(1998)}]{luchinsky1998}%
  \BibitemOpen
  \bibfield  {author} {\bibinfo {author} {\bibfnamefont {D.}~\bibnamefont
  {Luchinsky}}, \bibinfo {author} {\bibfnamefont {P.}~\bibnamefont
  {McClintock}}, \ and\ \bibinfo {author} {\bibfnamefont {M.}~\bibnamefont
  {Dykman}},\ }\bibfield  {title} {\enquote {\bibinfo {title} {Analogue studies
  of nonlinear systems},}\ }\href@noop {} {\bibfield  {journal} {\bibinfo
  {journal} {Reports on Progress in Physics}\ }\textbf {\bibinfo {volume}
  {61}},\ \bibinfo {pages} {889--997} (\bibinfo {year} {1998})}\BibitemShut
  {NoStop}%
\bibitem [{\citenamefont {Semenov}(2024)}]{semenov2024_book}%
  \BibitemOpen
  \bibfield  {author} {\bibinfo {author} {\bibfnamefont {V.}~\bibnamefont
  {Semenov}},\ }\href@noop {} {\emph {\bibinfo {title} {Electronic modelling of
  deterministic and stochastic oscillators}}}\ (\bibinfo  {publisher}
  {Springer},\ \bibinfo {year} {2024})\BibitemShut {NoStop}%
\end{thebibliography}

%merlin.mbs aipnum4-1.bst 2010-07-25 4.21a (PWD, AO, DPC) hacked
%Control: key (0)
%Control: author (8) initials jnrlst
%Control: editor formatted (1) identically to author
%Control: production of article title (0) allowed
%Control: page (1) range
%Control: year (1) truncated
%Control: production of eprint (0) enabled
%

\end{document}